\documentclass[slac_one]{revtex4}
\usepackage{graphicx}
\usepackage{fancyhdr}
\begin{document}

\title{{\small{Hadron Collider Physics Symposium (HCP2008),
Galena, Illinois, USA}}\\ %% Please keep this conference title here
\vspace{12pt}
Parton Distribution Functions: Impact of HERA} %% Paper title goes here


\author{K. Nagano\\
On behalf of the H1 and ZEUS Collaborations}
\affiliation{KEK, Tsukuba, Ibaraki 305-0081, Japan}

\begin{abstract}
Recent progresses of the proton structure
measurements and determination of parton distribution functions
by $ep$ collisions at HERA are introduced.
\end{abstract}

\maketitle

\thispagestyle{fancy}


%%%%%%%%%%%%%%%%%%%%%%%%%%%%%%%%%%%%%%%%%%%%%%%%%%%%%%%%%%%%%%%%%%%%%%%%%%%%%%%%%%%
\section{INTRODUCTION}
%%%%%%%%%%%%%%%%%%%%%%%%%%%%%%%%%%%%%%%%%%%%%%%%%%%%%%%%%%%%%%%%%%%%%%%%%%%%%%%%%%%

Deep inelastic scattering (DIS) of leptons off nucleons
has been the key for our understanding
of the structure of the nucleon.
HERA at DESY is a unique facility for colliding
electrons (or positrons) with protons.
Compared with previous fixed-target DIS
experiments, the large center-of-mass energy of about 320~GeV allows
an extension of the explorable kinematic phase space
by two orders of magnitude both in $Q^2$, the
negative of the four-momentum transfer squared,
and in the Bjorken scaling variable $x$.
The maximum $Q^2$ reaches almost
10$^5$~GeV$^2$, which corresponds to a spatial resolution
of $10^{-16}$~cm.
The minimum $x$, which means
the momentum fraction carried by the struck parton~\footnote{
In the proton's infinite momentum frame.
},
reaches almost $10^{-5}$.
The cross section of the neutral-current (NC) DIS interaction,
$e^{\pm} p \rightarrow e^{\pm} X$, can be written as
\begin{equation}
  \frac{d^2\sigma(e^{\pm}p \rightarrow e^{\pm}X)}{dxdQ^2}
        = \frac{2\pi{\alpha}^2}{xQ^4}
  \{ Y_{+} F_2 \mp Y_{-} x F_3 -  y^2 F_L \},
  \label{eq:nc_xsec}
\end{equation}
where $\alpha$ is the fine structure constant,
$Y_{\pm} = 1 \pm (1-y)^2$ with $y$ being the inelasticity,
and $F_2$, $x F_3$ and $F_L$ are the structure functions
of the proton.
In the framework of the perturbative QCD inspired
quark parton model, the structure functions can be directly related to
the parton distribution functions (PDFs)
which are probability densities of partons existing inside proton.
At low $Q^2$, dominantly contributing to the cross section
is $F_2$ which is an electric-charge squared weighted
sum of all flavor quark PDFs.
In the low-$x$ region, $F_2$ is dominated by sea-quark PDFs,
and the DGLAP evolution of QCD ascribes the $Q^2$-dependence of
$F_2$ (``scaling violation'') as largely owing
to gluon splitting into $q\overline{q}$-pairs.
Thus, HERA data provide crucial information on the small-$x$ sea-quark and gluon PDFs.
At large $Q^2$, $x F_3$ becomes significant, and
gives information on valence quark PDFs.
The structure function $F_L$ is zero in the naive quark-parton model,
i.e. without QCD, but in leading order QCD, a finite value of $F_L$ is expected
in the small $x$ region by being directly related to the gluon PDF.
The cross section of the charged-current (CC) DIS interaction,
$e^+ (e^-) p \rightarrow \overline{\nu} (\nu) X$, can be written as
\begin{eqnarray}
\frac{d^2\sigma(e^+p)}{dxdQ^2} &=& \frac{G_F^2}{2\pi}\frac{M_W^4}{(Q^2+M_W^2)^2}
         [ (\overline{u}+\overline{c}) + (1-y)^2 (d + s) ],\\
\frac{d^2\sigma(e^-p)}{dxdQ^2} &=& \frac{G_F^2}{2\pi}\frac{M_W^4}{(Q^2+M_W^2)^2}
       [ (u+c) + (1-y)^2 (\overline{d} + \overline{s}) ],
\end{eqnarray}
thus bringing flavor sensitivity of the valence quark PDFs at large $x$.

Based on about 100 (20) pb$^{-1}$ of
$e^+ p$ ($e^- p$) data collected
until the year 2000 (HERA-I), 
both the H1 and ZEUS experiments at HERA have measured 
the inclusive $e^{\pm}p$ NC and CC DIS cross sections~\cite{
        ref:h1_f2,
        ref:h1_lowq2,
        ref:h1_nccc_9899,
        ref:h1_nccc_9900,
        ref:zeus_f2,
        ref:zeus_cc_9497,
        ref:zeus_nc_9899,
        ref:zeus_cc_9899,
        ref:zeus_nc_9900,
        ref:zeus_cc_9900}.
After the year 2000, HERA underwent a major upgrade (HERA-II)
aiming for higher luminosity~\footnote{
The other aim of HERA-II is to provide collisions
with longitudinally polarized $e^{\pm}$ beams, giving
a direct sensitivity to the helicity structure of the
Electro-Weak interaction.},
and until March 2007, HERA provided about 500 pb$^{-1}$ of
$e^{\pm}p$ collisions to each H1 and ZEUS experiments.
The increased statistics provides more sensitivity to the PDFs at
large $x$ and large $Q^2$, in regions where the precisions of
the HERA-I measurements are still statistically limited.
In addition, the precision of the structure function measurements by tagging
heavy-quarks in the final state can be significantly 
improved by making use of the increased statistics
of HERA-II and also increased capability of
heavy-quark identification with newly installed
tracking devices.
During March to June in the year 2007, HERA made
a series of dedicated runs with 
reduced proton beam energies of 460 and 575~GeV as compared
to the nominal one of 920~GeV.
These data sets are essential in the first direct measurement 
of $F_L$, as will be discussed later.
Although HERA ended its 15 years spanning operation history 
in June 2007, analyses using full data sets are
lively ongoing, and significant progresses are being made.
In this manuscript, recent results from HERA 
are presented, and their impacts are discussed.

%%%%%%%%%%%%%%%%%%%%%%%%%%%%%%%%%%%%%%%%%%%%%%%%%%%%%%%%%%%%%%%%%%%%%%%%%%%%%%%%%%%
\section{H1 AND ZEUS COMBINED ANALYSIS}
%%%%%%%%%%%%%%%%%%%%%%%%%%%%%%%%%%%%%%%%%%%%%%%%%%%%%%%%%%%%%%%%%%%%%%%%%%%%%%%%%%%

\subsection{Combined cross sections}

PDFs are usually determined in global QCD analyses of
DIS data, both from HERA and fixed-target experiments
as well as jet production data from the TEVATRON~\cite{ref:mrst, ref:cteq}.
Given that high-$Q^2$ HERA data
which are sensitive to large $x$ PDFs are now available,
H1 and ZEUS used their own data alone to make PDF fits
(called H1 PDF 2000 and ZEUS-JETS fits, respectively)~\cite{ref:h1pdf2000,ref:zeus_jets}.
Statistical uncertainties
of these data are very small at low $Q^2$ such that
total uncertainties
are dominated by systematic uncertainties, and
data points are correlated through
common systematic uncertainties, both within and
across the data sets. Thus, consideration
of point-to-point correlations between systematic
uncertainties was essential.
In fact, this issue is faced by all of such QCD analyses;
there are basically two methods, 
the Hessian and Offset methods, but each method has
its own advantages and shortcomings, and there is
no agreed standard (see e.g.~\cite{ref:zeus_s}).
For example, in the Hessian method, each systematic error source is
treated as an additional fit parameter
which is optimized by the model provided by QCD.
However, difficulties arise when there are
inconsistencies between data sets,
leading to shifts of systematic error parameters
by far more than one standard deviation~\footnote{
Some global QCD analyses thus use non-statistical criteria to
estimate the PDF uncertainties ($\delta\chi^2=T^2 >> 1)$.
}.

These drawbacks can be significantly
reduced by averaging the data sets
in a model-independent way prior to performing
a QCD analysis on them.
Recently, the H1 and ZEUS experiments have 
made a combined analysis on their data
in order to obtain 
such averaged data sets~\cite{ref:combxs}.
Almost all inclusive DIS cross section measurements
with HERA-I data, including low $Q^2$ NC as well as high $Q^2$ NC and CC,
are combined~\cite{
        ref:h1_f2,
        ref:h1_lowq2,
        ref:h1_nccc_9899,
        ref:h1_nccc_9900,
        ref:zeus_f2,
        ref:zeus_cc_9497,
        ref:zeus_nc_9899,
        ref:zeus_cc_9899,
        ref:zeus_nc_9900,
        ref:zeus_cc_9900}.
Since both sets of cross section data from H1 and ZEUS
measure the same true cross-sections, 
the combination procedure is a $\chi^2$-minimization
in which the parameters are the true values of the
cross-section at each ($x$, $Q^2$) point
and the correlated systematic error parameters.
These parameters are thus determined in a
model independent way using the Hessian method~\cite{ref:sasha_average}.
This combination procedure cross-calibrates the measurements,
such that the overall uncertainty is reduced
by much more than the simple average.
In total, 1153 individual NC and CC
measurements are averaged to 554 unique points,
and 43 sources of correlated systematic uncertainty parameters
are fitted.
This yields a good quality of fit with the $\chi^2/dof$=510/599.
In the small $Q^2$ region of $Q^2 < 12$~GeV$^2$ the combined data
set has precision much better than 2\%, whereas H1 and ZEUS
data each have a precision of $\sim 3$~\%.
\begin{figure*}[t]
\centering
\includegraphics[width=0.47\textwidth]{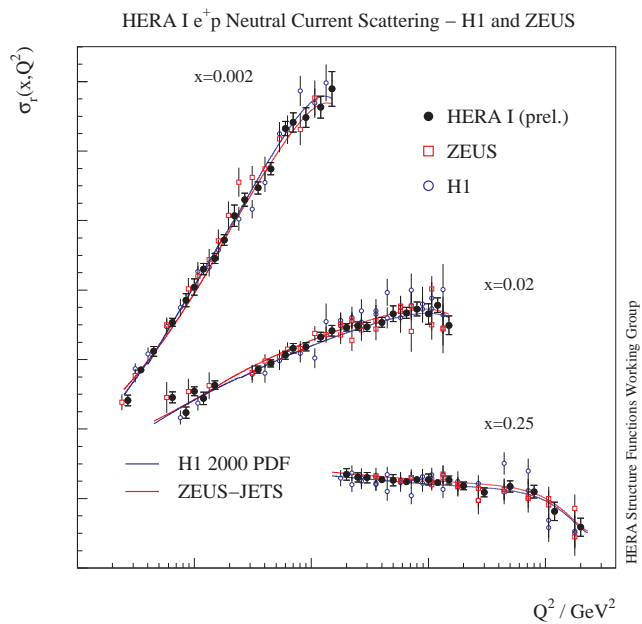}
\caption{H1 and ZEUS combined cross sections
for three selected $x$ bins as a function $Q^2$,
together with the H1 PDF 2000 and ZEUS-JETS predictions.
}
\label{fig:f2ave}
\end{figure*}
Figure~\ref{fig:f2ave} illustrates the H1 and ZEUS combined cross
sections for three selected $x$ bins as a function of $Q^2$.
The H1 (open circles) and ZEUS data (open squared) are
compared to the H1 and ZEUS combined data (closed points).
Measurements from the individual experiments
have been displaced sideways for clarity.
The error bars show the total uncertainty.
At low $Q^2$ where the data are limited by
systematic uncertainties, improvement in the total
error is impressive.
At higher $Q^2$ the combined data exhibit far smaller
fluctuations as compared to individual data
due to the increased statistical accuracy.

\subsection{Combined PDF: HERAPDF}

A  next-to-leading order (NLO) QCD analysis
was performed using 
the combined HERA-I data set of $e^{\pm}p$ NC and CC cross sections
presented in the previous subsection as the sole input
(called HERAPDF0.1)~\cite{ref:combfit}.
The PDFs are parameterized at $Q^2_0 = 4$~GeV$^2$
with 11 free parameters.
The constraints of the number sum-rules and the momentum sum rule
are applied.
All the data have values of the invariant mass of the
hadronic system, $W^2 > 300$~GeV$^2$
and of $x < 0.65$ such that there is no need to account for
target mass corrections or higher twist effects.
Also, a cut of $Q^2 > 3.5$~GeV$^2$ is applied to 
avoid non perturbative effects.
As the correlated systematic uncertainties are no longer
crucial in the HERA-I combined data sets,
different handling of systematic uncertainties resulted
in similar fit results.
For the central fit, the 43 systematic uncertainties from the
individual H1 and ZEUS data sets are quadratically combined
to the statistical uncertainties, and the 4 sources of
uncertainty from the combination procedure
is treated with the Offset method.
The $\chi^2$ per degree of freedom of the fit was 477/562.
\begin{figure*}[htb]
\centering
\includegraphics[width=0.38\textwidth]{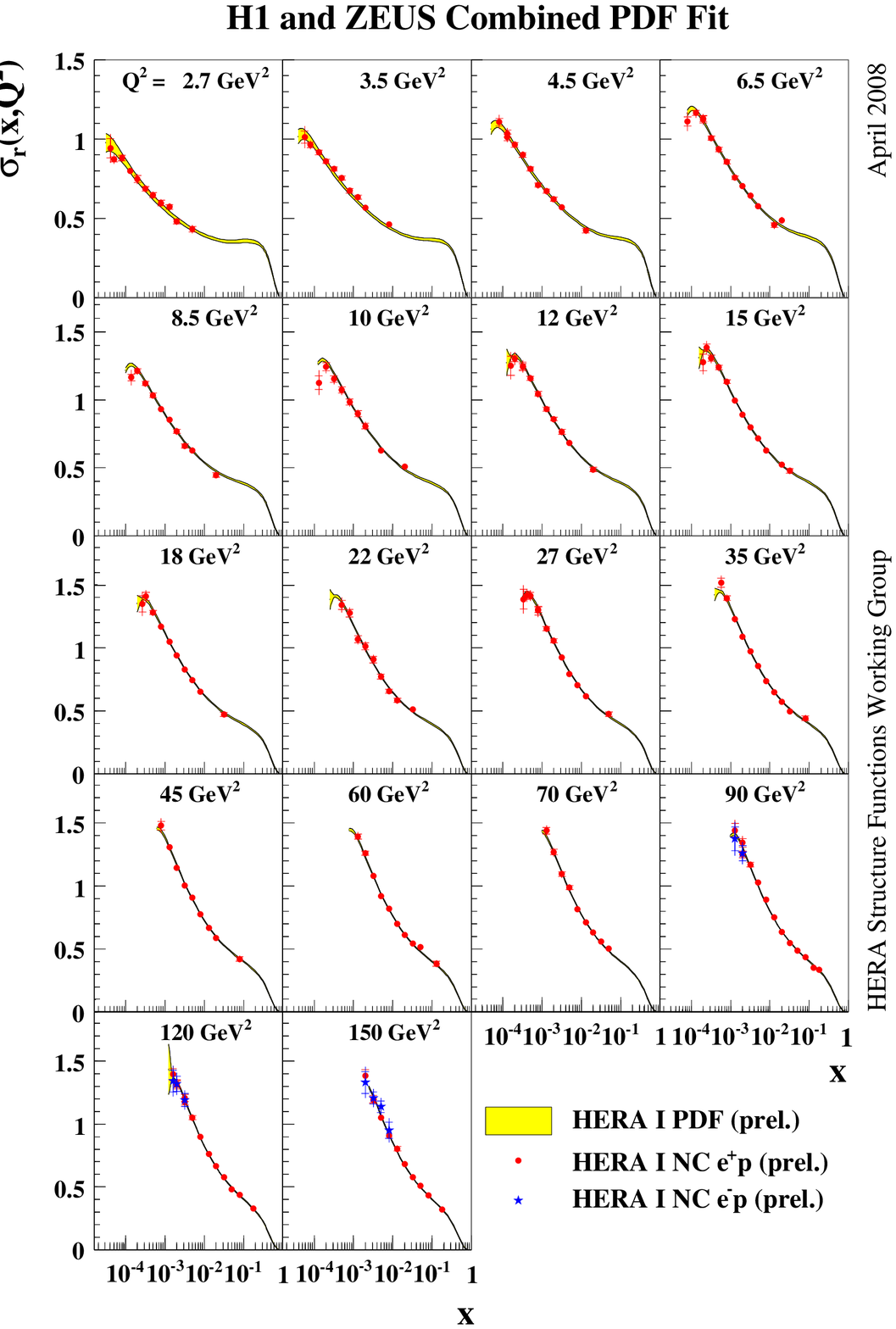}
\hspace{5mm}
\includegraphics[width=0.38\textwidth]{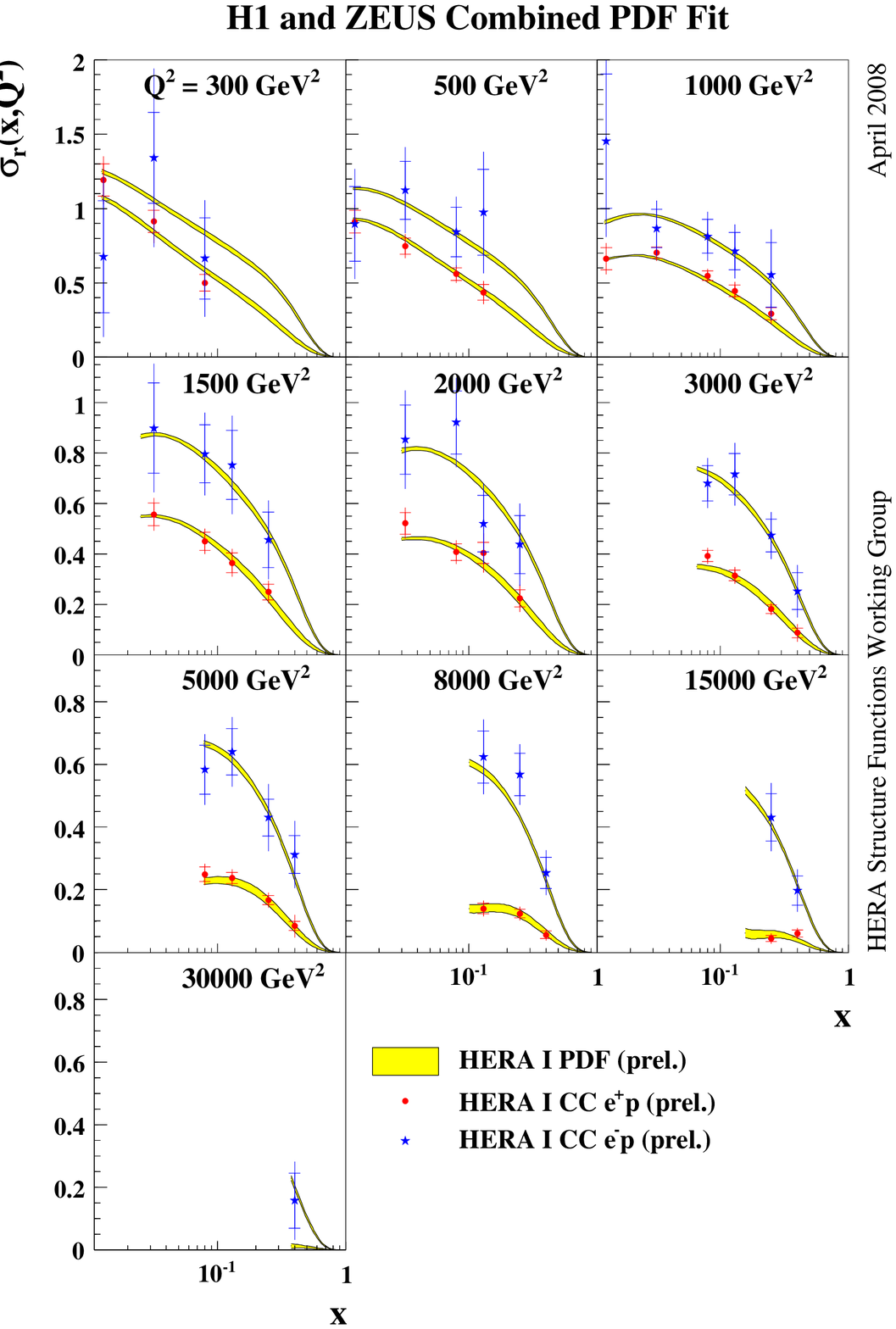}
\caption{Combined HERA-I $e^{\pm}p$ 
NC (left) and CC (right) cross sections, together with the fit results of HERAPDF0.1.}
\label{fig:combfit_nccc}
\end{figure*}
\begin{figure*}[htb]
\centering
\includegraphics[width=0.38\textwidth]{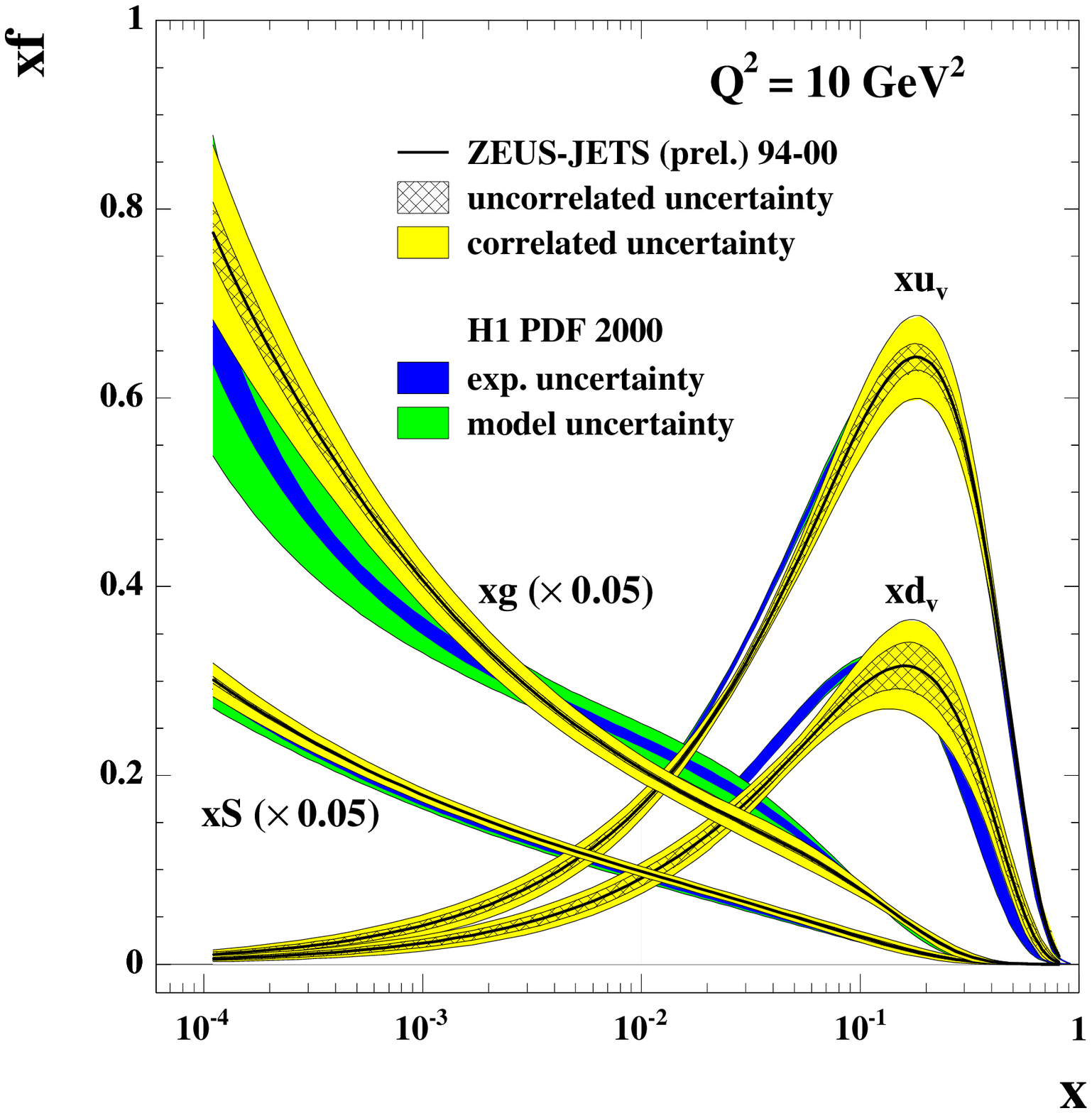}
\hspace{5mm}
\includegraphics[width=0.38\textwidth]{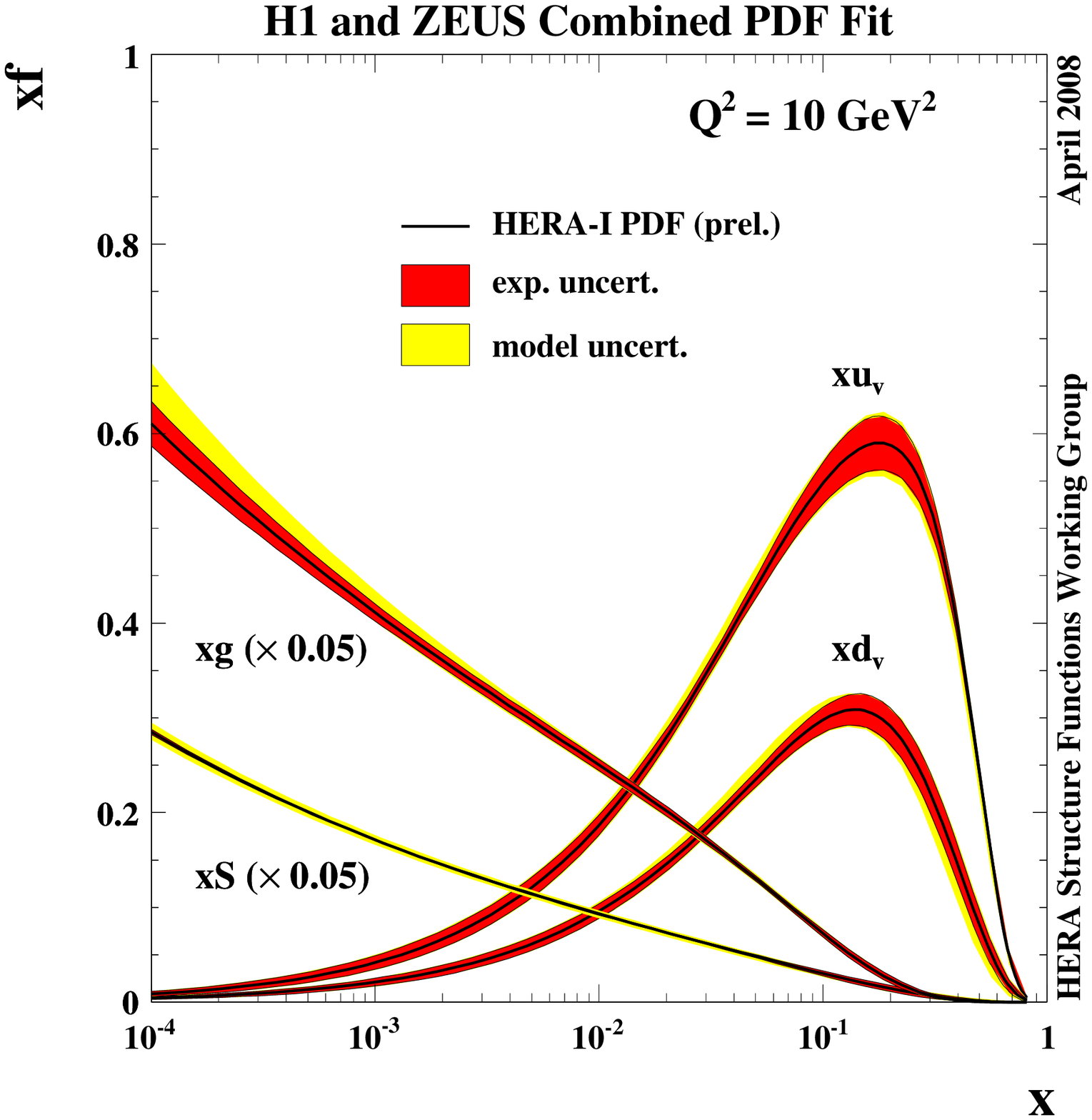}
\caption{H1 PDF 2000 and ZEUS-JETS PDFs (left) and HERAPDF0.1
PDFs (right) at $Q^2=10$~GeV$^2$.}
\label{fig:combfit_pdf}
\end{figure*}

Figure~\ref{fig:combfit_nccc} shows 
the combined HERA-I cross sections together with
the HERAPDF0.1 fit results.
The fit gives a good description both to NC and CC
cross sections, and both for $e^{\pm}$ collisions.
Figure~\ref{fig:combfit_pdf} shows 
the HERAPDF0.1 at $Q^2$=10~GeV$^2$ (right plot).
Plotted together in the left plot 
are the H1 2000 PDF and ZEUS-JET PDFs, i.e.
obtained by H1 and ZEUS fitting to their
own data.
They have rather different shapes,
and the size of the uncertainties are compatible to those of
the global PDF analyses.
This clearly shows that HERAPDF0.1 has a much better
determination benefiting from the high precision of
the combined data sets in which inconsistencies between the two data sets 
are resolved by the cross calibration.
It is worth noting that the large amount of HERA-II data 
are not yet included in the combined data sets,
and thus further improvements are foreseen
for valence quark PDFs at large $x$, in particular 
for the $d$-quark PDF which will be determined from $e^+p$ CC data.
\begin{figure*}[htb]
\centering
\includegraphics[width=0.38\textwidth]{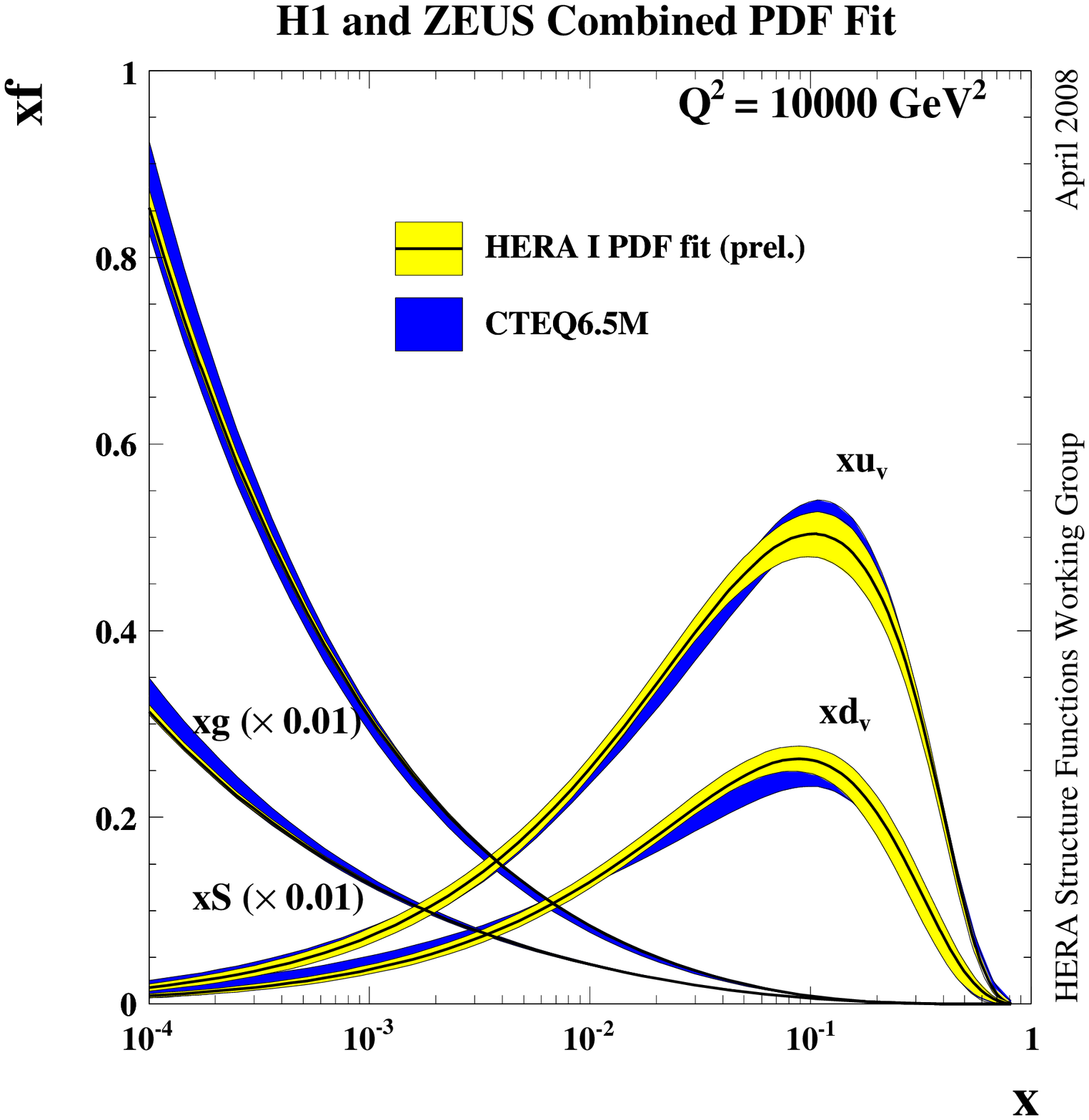}
\caption{HERAPDF0.1 at $Q^2=10000$~GeV$^2$ compared to CTEQ6.5 PDF.}
\label{fig:combfit_pdf_hq2}
\end{figure*}
Figure~\ref{fig:combfit_pdf_hq2} presents the HERAPDF0.1 at large $Q^2$ of
10000~GeV$^2$ compared to one of the global PDF analyses
(CTEQ 6.5~\cite{ref:cteq} is chosen here), showing that  
the improvement in precision of PDFs holds until large $Q^2$.
The HERAPDF0.1 has a significant impact on the precision of the 
predictions for Standard Model processes such as $W$ and $Z$ production
at LHC~\cite{ref:mandy_dis07proc}.
This due to the improved precision of 
the sea and the gluon PDFs at $Q^2 \sim 10000$~GeV$^2$
and $5\times10^{-4} < x < 5\times10^{-2}$.
In fact, it has been shown that the precision of the predicted
cross-sections at LHC is less than 2\% in the mid-rapidity range for HERAPDF0.1
PDFs as compared to 15\% for pre-HERA PDFs~\cite{ref:mandy_pdf4lhc}.

%%%%%%%%%%%%%%%%%%%%%%%%%%%%%%%%%%%%%%%%%%%%%%%%%%%%%%%%%%%%%%%%%%%%%%%%%%%%%%%%%%%
\section{DIRECT MEASUREMENT OF \boldmath{$F_L$}}
\label{sect:fl}
%%%%%%%%%%%%%%%%%%%%%%%%%%%%%%%%%%%%%%%%%%%%%%%%%%%%%%%%%%%%%%%%%%%%%%%%%%%%%%%%%%%

Despite of the remarkable success of QCD
as seen in previous sections,
there are still several issues to be clarified.
For example, validity of
the DGLAP formalism should be confirmed more explicitly at small $x$.
This is because the DGLAP equation is a leading log re-summation
of $\ln Q^2$ terms, and thus it cannot be
expected to work at small $x$ where large $\ln (1/x)$ terms will emerge.
Notice also that the gluon PDF is extracted 'indirectly'
through the $Q^2$-dependence of inclusive cross sections,
and thus should be further extensively 
verified with other processes which have more direct sensitivity.
In both of these senses, a measurement of $F_L$
is of importance.
As $F_L$ is zero in a static view of proton,
it is related to the dynamical picture of the proton.
Therefore, $F_L$ is in particular
interesting at small $x$, and only HERA can perform such measurement.
In the DGLAP formalism $F_L$ is related to the gluon PDF at small $x$,
thus giving a confirmation of indirectly determined
gluon PDF in QCD analyses.

The structure function $F_L$ can be extracted from sets of
cross sections measured with different center-of-mass energies, $s$,
as follows:
if we present cross sections in a reduced form $\sigma_r$ defined as
\begin{equation}
\sigma_r \equiv \frac{d^2\sigma}{dxdQ^2} \frac{xQ^4}{2\pi\alpha^2 Y_+} 
= F_2 - \frac{y^2}{Y_+}F_L,
\end{equation}
$F_L$ can be extracted from measurements of $\sigma_r$ made at
various $s$, from the slope of $\sigma_r$ with $y^2 / Y_+$~\footnote{
Contribution from $xF_3$ can be neglected at small $Q^2$. 
}.
Note that for measurements at fixed ($x$, $Q^2$) variation of $y$
requires variation of $s$, since $Q^2=xys$.
The precision of the extracted $F_L$ is dominated by 
the size of the 'lever arm' spanned in the axis of
$y^2 / Y_+$, and thus, extending the measurement up to $y$ as high as possible is crucial.
At low $Q^2$, high $y$ values correspond to low values of
the scattered electron energy.
As small energy depositions can also be caused by
hadronic final state particles and hence fake electrons,
there are large backgrounds expected at high $y$
dominantly due to photoproduction precesses at $Q^2 \approx 0$~GeV$^2$.
In HERA-II, H1 and ZEUS has developed new measurement techniques
optimized for large $y$. These were then validated with
nominal beam energy data~\cite{ref:h1_highy, ref:zeus_highy}
before starting the dedicated runs with reduced proton energies.
In the H1 analysis, photoproduction backgrounds are statistically
subtracted by using the sample in which wrong sign of charge
is reconstructed for the track linked to the electron candidates
measured in calorimeter. 
A small correction to charge asymmetry in 
backgrounds was obtained from comparisons
of samples of negative tracks in $e^+p$ scattering with
samples of positive tracks in $e^-p$ scattering.
In the ZEUS analysis, a part of the photoproduction backgrounds are
tagged by using a special detector located close to the beampipe,
and are used to understand and normalize the background MC samples.
These experimental techniques are used in the 
analyses of special runs with reduced proton beam energies, i.e.
in the $F_L$ measurements.
H1 has measured $F_L$ in middle to high $Q^2$ region,
$12 < Q^2 < 800$~GeV$^2$~\cite{ref:h1_fl_midq2,ref:h1_fl_highq2}.
\begin{figure*}[tb]
\centering
\includegraphics[width=0.45\textwidth]{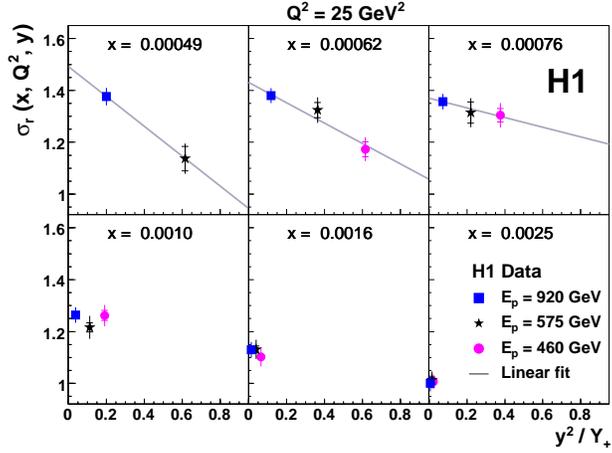}
\caption{Reduced cross section $\sigma_r$ at $Q^2=25$~GeV$^2$
as a function of $y^2/Y_+$, as measured by H1.}
\label{fig:fl_3fit}
\end{figure*}
Figure~\ref{fig:fl_3fit} shows the reduced cross section measured
with the three different proton beam energies, plotted as function
of $y^2/Y_+$. 
The data clearly shows a finite value of slope, hence $F_L$.
\begin{figure*}[tb]
\centering
\includegraphics[width=0.65\textwidth]{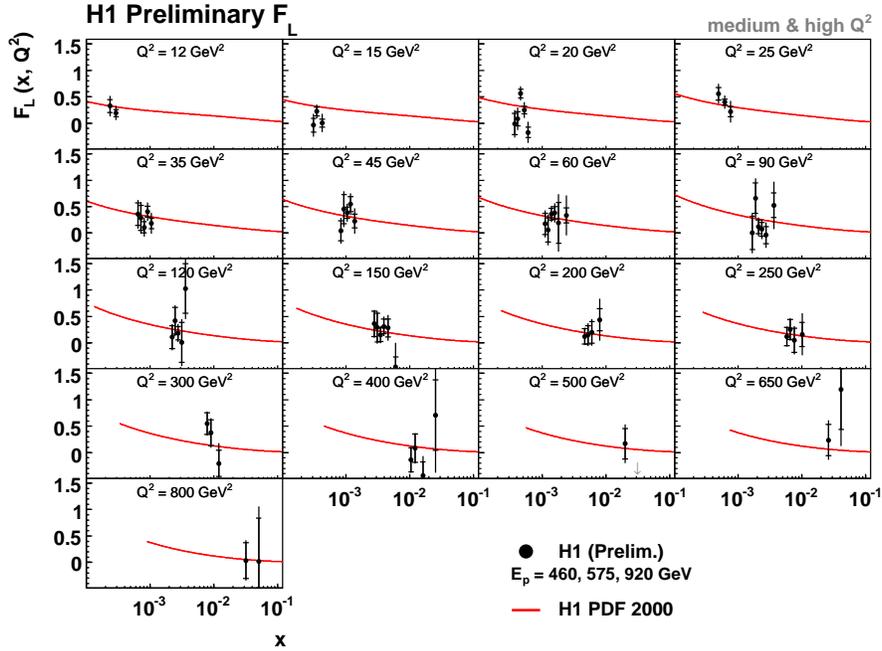}
\caption{$F_L$ as a function of $x$ at various $Q^2$, as measured by H1.}
\label{fig:fl_x}
\end{figure*}
\begin{figure*}[tb]
\centering
\includegraphics[width=0.45\textwidth]{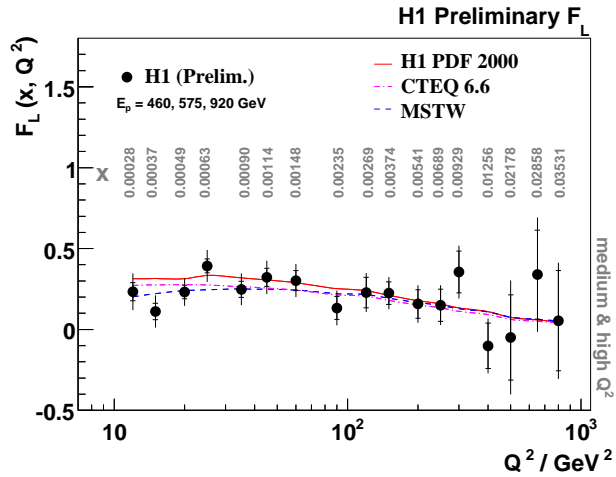}
\caption{$F_L$ as a function of $Q^2$ at given values of $x$,
compared with QCD predictions.}
\label{fig:fl_q2}
\end{figure*}
Figure~\ref{fig:fl_x} shows the extracted $F_L$ as a function of $x$
in various $Q^2$ bins. 
The result is consistent with the prediction obtained with the
H1 PDF 2000 fit.
The values on $F_L$ resulting from averages over $x$ at fixed
$Q^2$ are shown in Figure~\ref{fig:fl_q2}.
Within the experimental uncertainties, the data are consistent with the
QCD predictions, indicating the applicability of the DGLAP formalism at small
$x$ at HERA.
\begin{figure*}[htb]
\centering
\includegraphics[width=0.45\textwidth]{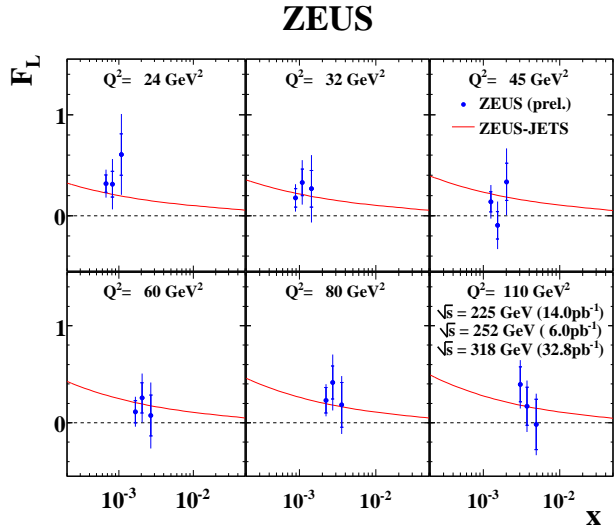}
\caption{$F_L$ as a function of $x$ at various $Q^2$, as measured by ZEUS.}
\label{fig:zeus_fl}
\end{figure*}
While, Figure~\ref{fig:zeus_fl} shows the $F_L$ as measured by ZEUS~\cite{ref:zeus_fl}.
The measurement is consistent with the
ZEUS-JETS prediction of $F_L$, i.e. a QCD prediction, but also with $F_L=0$,
within the current precision of the measurement.

%%%%%%%%%%%%%%%%%%%%%%%%%%%%%%%%%%%%%%%%%%%%%%%%%%%%%%%%%%%%%%%%%%%%%%%%%%%%%%%%%%%
\section{HEAVY QUARK CONTRIBUTION TO \boldmath{$F_2$}}
%%%%%%%%%%%%%%%%%%%%%%%%%%%%%%%%%%%%%%%%%%%%%%%%%%%%%%%%%%%%%%%%%%%%%%%%%%%%%%%%%%%

In leading order of QCD, the heavy flavor production in $ep$ collisions
is dominated by the photon gluon fusion
process (BGF), $\gamma g \rightarrow Q\overline{Q}$.
Hence, it is sensitive to the gluon PDF
and allows tests of the universality of the gluon PDF.
The heavy quark contribution to $F_2$, $F_2^{Q\overline{Q}}$, 
can be obtained by extrapolating the visible
measured cross section of heavy quark production to the full phase space
using NLO QCD calculations.
\begin{figure*}[htb]
\centering
\includegraphics[width=0.38\textwidth]{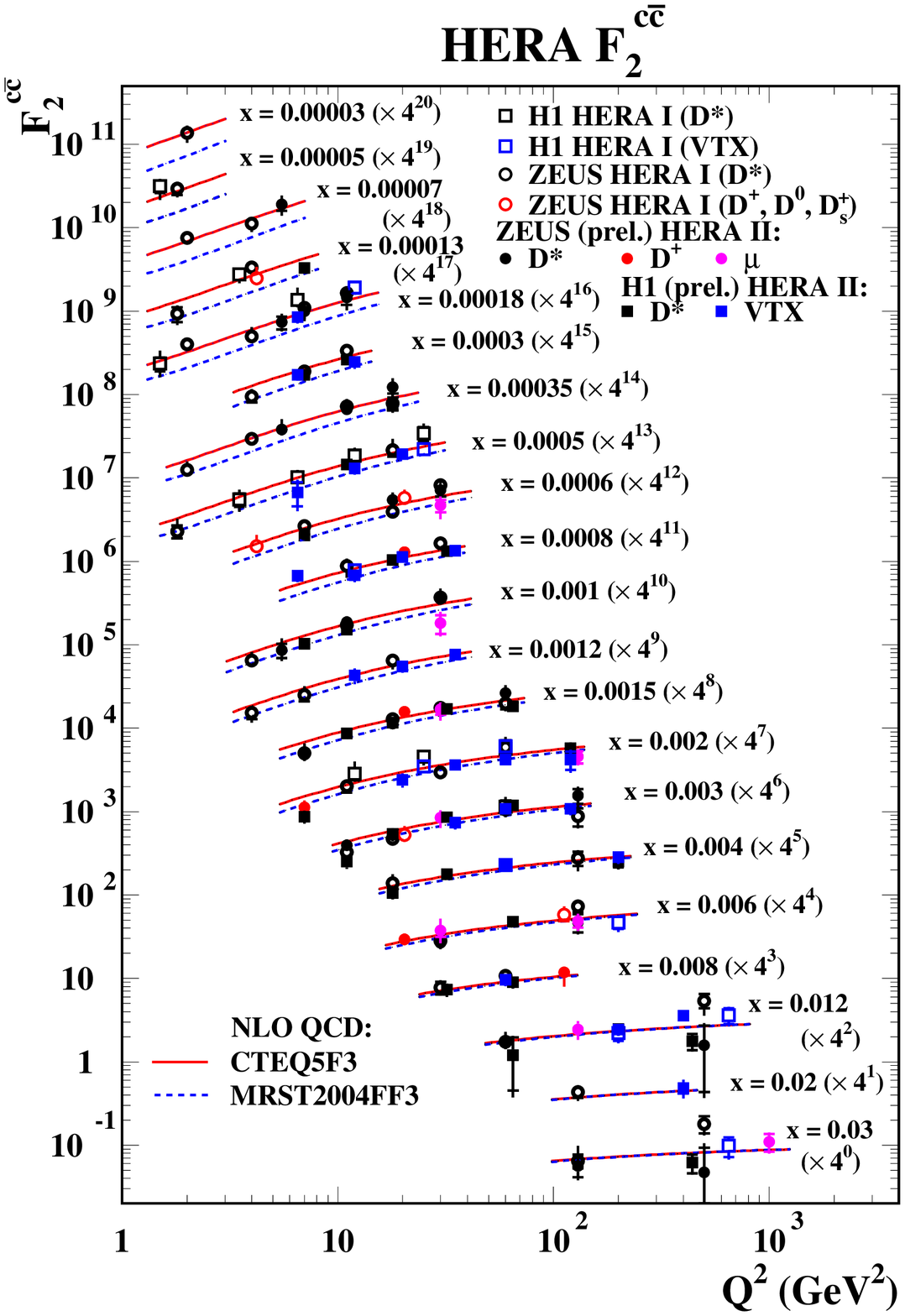}
\hspace{5mm}
\includegraphics[width=0.38\textwidth]{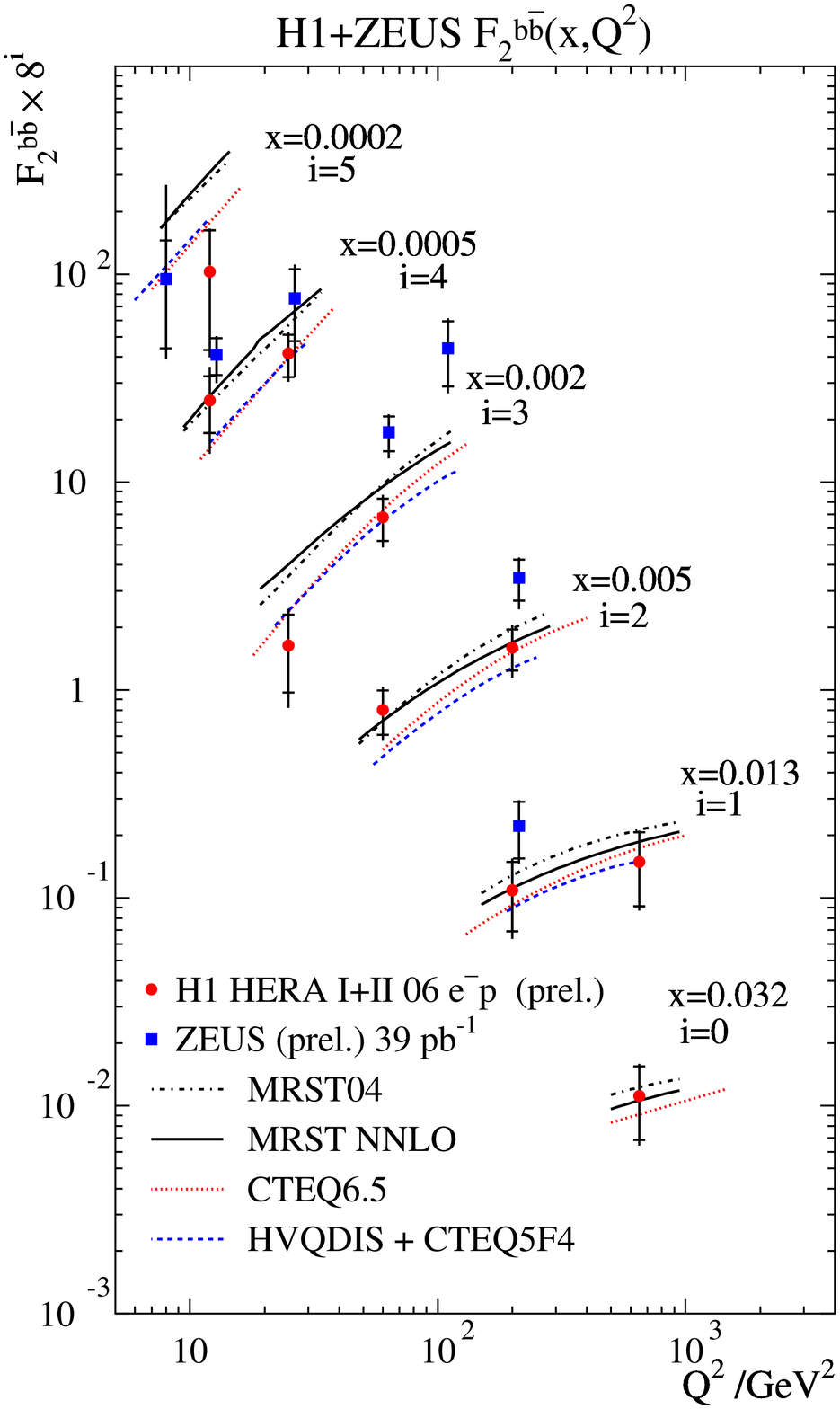}
\caption{Heavy quark contribution to the structure function $F_2$: 
charm contribution $F_2^{c\overline{c}}$ (left) and
bottom contribution $F_2^{b\overline{b}}$ (right).}
\label{fig:f2_hq}
\end{figure*}

Figure~\ref{fig:f2_hq} shows the 
charm contribution to $F_2$, $F_2^{c\overline{c}}$,
as measured by H1 and ZEUS~\cite{
        ref:zeus_f2c_dstar_hera2,
        ref:zeus_f2c_d0_dcharged_hera2,
        ref:zeus_f2c_dmesons_hera1,
        ref:h1_f2c_dstar_hera2,
        ref:h1_f2c_dstar_hera1,
        ref:h1_f2c_f2b_vtx_hera2,
        ref:h1_f2c_f2b_vtx_hera1_lowq2,
        ref:h1_f2c_f2b_vtx_hera1_highq2,
        ref:zeus_f2b_hera2}.
By using about 160~pb$^{-1}$ of luminosity of HERA-II data,
ZEUS has measured the $D^{*\pm}$ meson production cross section
by tagging it with slow $\pi$~\cite{ref:zeus_f2c_dstar_hera2}.
Production of $D^{\pm}$ and $D^{0}$ mesons have also been measured
making use of the capability of secondary vertex identification
of the Micro Vertex Detector (MVD) which was installed
at HERA-II~\cite{ref:zeus_f2c_d0_dcharged_hera2}.
A signed 2-dimensional decay length significance with respect
to the beamspot measured in MVD gives a significant improvement
in the signal to noise ratio, and thus in the statistical uncertainty
of the measured cross section.
These analyses are based on HERA-II data of about 130 pb$^{-1}$ of luminosity.
The results are plotted in the left plot of Figure~\ref{fig:f2_hq},
together with previous
measurements by using HERA-I data, i.e. without
the advantage of the MVD~\cite{ref:zeus_f2c_dmesons_hera1}.
H1 has measured $D^{*\pm}$ meson production~\cite{ref:h1_f2c_dstar_hera2},
as well as the
inclusive charm and beauty production~\cite{ref:h1_f2c_f2b_vtx_hera2} cross sections,
based on HERA-II data of about 350 and 190~pb$^{-1}$ of luminosities, respectively.
In the inclusive analysis, 
the charm and beauty fractions are determined
using a neural network which includes the impact
parameter of tracks to the primary vertex
and the position of the secondary vertex
as measured using the Vertex Detector (VTX).
These results are also shown in the left plot of Figure~\ref{fig:f2_hq},
compared to previous measurements
using HERA-I data set~\cite{
ref:h1_f2c_dstar_hera1,
ref:h1_f2c_f2b_vtx_hera1_lowq2,
ref:h1_f2c_f2b_vtx_hera1_highq2}.
The $F_2^{c\overline{c}}$ measurement
using the VTX gives a consistent
result with that measured
using the technique of tagging $D^{*\pm}$ meson,
thus providing a proof of the VTX analysis.

In the right plot of Figure~\ref{fig:f2_hq} shown is
the bottom-quark contribution to $F_2$, $F_2^{b\overline{b}}$,
as measured by H1 in the inclusive analysis using VTX~\cite{
        ref:h1_f2c_f2b_vtx_hera2,
        ref:h1_f2c_f2b_vtx_hera1_lowq2,
        ref:h1_f2c_f2b_vtx_hera1_highq2}.
Also plotted is the ZEUS measurement tagging bottom quarks
by requiring muon associated with jet, based on
HERA-II data sample of about 40~pb$^{-1}$~\cite{ref:zeus_f2b_hera2}.
In these $F_2^{c\overline{c}}$ and $F_2^{b\overline{b}}$
measurements, 
it is shown that
a significant fraction of $F_2$ is owed by heavy quarks,
about 30\% by charm and about 3\% by bottom,
increasing with $Q^2$.
Precision of determining heavy-quark PDFs will 
be further improved by the analyses with full HERA-II data.

%%%%%%%%%%%%%%%%%%%%%%%%%%%%%%%%%%%%%%%%%%%%%%%%%%%%%%%%%%%%%%%%%%%%%%%%%%%%%%%%%%%
\section{MEASUREMENT OF \boldmath{$xF_3$}}
%%%%%%%%%%%%%%%%%%%%%%%%%%%%%%%%%%%%%%%%%%%%%%%%%%%%%%%%%%%%%%%%%%%%%%%%%%%%%%%%%%%

The difference of cross sections between $e^+p$ and $e^-p$ collision data
determines $xF_3$ as seen in Eq.~\ref{eq:nc_xsec}.
At HERA, $xF_3$ is, to a very good approximation, dominated 
only by the $\gamma$-$Z$ interference term, $xF_3^{\gamma Z}$, as
\begin{equation}
\sigma_r(e^-) - \sigma_r(e^+) = - 2 \frac{Y_-}{Y_+}
 a_e k_Z xF_3^{\gamma Z},
\end{equation}
where $\sigma_r$ is the reduced cross section
defined in Section~\ref{sect:fl},
$k_Z = \frac{1}{4\sin^2\theta_W\cos^2\theta_W}
\frac{Q^2}{Q^2+M_Z^2}$
with $M_Z$ being the mass of $Z$ boson
and $\theta_W$ being the Weinberg angle,
and $a_e=-1/2$ is the axial-vector coupling of
electron to the $Z$ boson.
In leading order QCD, $xF_3^{\gamma Z}$ can be written as
\begin{equation}
xF_3^{\gamma Z} = 2 x [ e_u a_u u_V + e_d a_d d_V],
\end{equation}
with an assumption that 
$\Delta_u = (u_{sea} - \overline{u} + c - \overline{c})$
and 
$\Delta_d = (d_{sea} - \overline{d} + s - \overline{s})$
are zero,
where $e_u$ and $a_u$ are the charge and axial-vector couplings to the
$Z$ of the $u$-quark, and $e_d$ and $a_d$ are the corresponding
quantities for the $d$-quark.
A sum rule holds in leading order as
\begin{equation}
\int_{0}^{1} xF_3^{\gamma Z} \frac{dx}{x} = \frac{1}{3}\int_{0}^{1}
        (2 u_V + d_V ) dx = \frac{5}{3}.
\end{equation}
The structure function $xF_3^{\gamma Z}$ is thus
determined by the valence quark PDFs and predicted
to be only very weakly depending on $Q^2$.
\begin{figure*}[tb]
\centering
\includegraphics[width=0.38\textwidth]{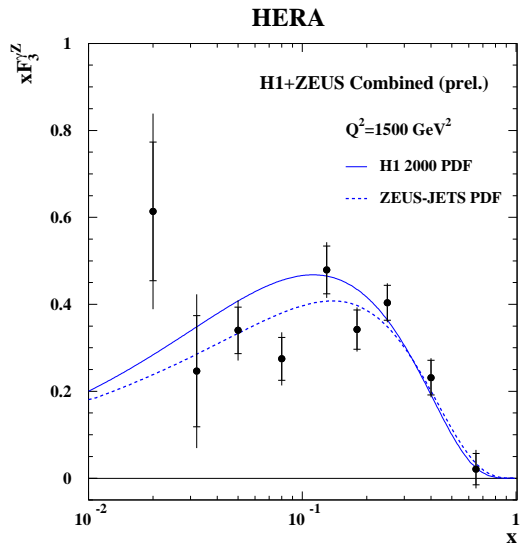}
\caption{The $\gamma$-$Z$ interference
term of the structure function $xF_3$, $xF_3^{\gamma Z}$.}
\label{fig:xf3gz}
\end{figure*}

Figure~\ref{fig:xf3gz} shows the $xF_3^{\gamma Z}$ measured
by using both H1 and ZEUS high $Q^2$ 
NC cross sections from HERA-I and HERA-II~\cite{ref:xf3gz}.
The measurement is well described by 
the H1 2000 PDF and ZEUS-JETS fits.
The small difference between the H1 and ZEUS fits
indicates weak constraints on the behavior of the valence quark PDFs at smaller $x$.
In the range of acceptance, the integral of $F_3^{\gamma Z}$
is measured to be
\begin{equation}
\int_{0.02}^{0.65} F_3^{\gamma Z} = 1.21 \pm 0.09 (stat) \pm 0.08 (syst),
\end{equation}
which is consistent with the results of the H1 and ZEUS
QCD fits, $1.12 \pm 0.02$ and $1.06 \pm 0.02$, respectively.

%%%%%%%%%%%%%%%%%%%%%%%%%%%%%%%%%%%%%%%%%%%%%%%%%%%%%%%%%%%%%%%%%%%%%%%%%%%%%%%%%%%
\section{SUMMARY}
%%%%%%%%%%%%%%%%%%%%%%%%%%%%%%%%%%%%%%%%%%%%%%%%%%%%%%%%%%%%%%%%%%%%%%%%%%%%%%%%%%%

HERA has provided the most precise inclusive 
structure function measurements,
which has brought significant improvements to our knowledge on
proton structure.
For further comprehensive understanding,
analyses with new techniques and with the full amount of
data are being performed, and new high precision
results are being produced.
Final publications with ultimate precision will come
in the next years.

% \begin{acknowledgments}
% The authors wish to thank JACoW for their guidance in preparing
% this template.
% Work supported by Department of Energy contract DE-AC02-76SF00515.
% \end{acknowledgments}

%\begin{thebibliography}{9}   % Use for  1-9  references

\end{document}